\documentclass[prd,preprint,superscriptaddress]{iopart}
\usepackage{graphicx}
\usepackage{iopams}
\newcommand{\be}{\begin{equation}}
\newcommand{\ee}{\end{equation}}
\newcommand{\ben}{\begin{eqnarray}}
\newcommand{\een}{\end{eqnarray}}
\begin{document}

\title[Accelerated expansion of a universe containing a self-interacting BE gas]
{Accelerated expansion of a universe containing a self-interacting
Bose-Einstein gas}
\author{Germ\'{a}n Izquierdo \footnote{ E-mail address:
german.izquierdo@gmail.com}}
\author{Jaime Besprosvany \footnote{
E-mail address: bespro@fisica.unam.mx}}

\address{Instituto de F\'isica, Universidad Nacional Aut\'onoma de
M\'exico, Circuito de la Investigaci\'on Cient\'ifica S/N, Ciudad
Universitaria, CP 04510, M\'exico, Distrito Federal, M\'exico.}

\begin{abstract}
Acceleration of the universe is obtained from a model of
non-relativistic particles with a short-range attractive
interaction, at low enough temperature to produce a Bose-Einstein
condensate. Conditions are derived for negative-pressure behavior.
In particular, we show that a phantom-accelerated regime at the
beginning of the universe solves the horizon problem, consistently
with nucleosynthesis.
\end{abstract}
\pacs{98.80.-c, 98.80.cq, 95.36.+x} \maketitle
\section{Introduction}

There is observational evidence that the Universe is undergoing an
accelerated expansion driven by a cosmological constant or some
form of energy that violates the strong energy condition ($\rho+
3p>0$, $\rho$ and $p$ being the energy density and pressure,
respectively, of the fluid conforming the Universe)
\cite{consensus}. The observational data seem to admit models that
even assume that the Universe is dominated by a kind of fluid that
violates the dominant energy condition ($\rho+ p>0$). This fluid
is known in the literature as the phantom energy \cite{caldwell}.

Many types of models have been proposed in recent years to explain
the observed accelerated expansion: the cosmological constant,
which presents a fine-tuning problem \cite{coscons}; quintessence
models \cite{quint} and quintessential inflation fields
\cite{quintinfl}; scalar-field models \cite{scalfield}; chameleon
fields \cite{cham}; K-essence models \cite{Kesse}; modified
gravity models \cite{modgrav}; the feedback of non-linearities
into the evolution equations \cite{backreac}; Chaplygin gases
\cite{chapg}; tachyons \cite{tach};  phantom dark energy and ghost
condensates \cite{caldwell}; de-Sitter vacua with the flux
compactification in string theory \cite{vacc}; cyclic universe
\cite{cycl}. A more complete list and discussion can be found in
Ref. \cite{copeland}. Although all the models predict an
accelerated expansion that is in good agreement with the
observational data, few of them propose a microscopic
understanding of dark energy.

Another stage of accelerated expansion has been proposed over the
years in an attempt to explain the horizon, flatness and unwanted
relic problems: inflation \cite{infl, Kolb, Liddle}. Most
inflationary models assume that the expansion of the universe is
dominated by a scalar field with a slow-roll potential, so that
the scale factor evolves nearly exponentially with time. During
the inflationary stage of the expansion the scale factor increases
several orders of magnitude and the relic radiation energy density
decreases dramatically. To allow nucleosynthesis to happen in the
inflationary scenario, a reheating process should be introduced:
the inflaton field decays into conventional fields by means of the
potential. After reheating, the universe is dominated by
radiation.

Given the lack of information about the nature of the dark energy,
thermodynamics and holography may be helpful. As examples of this
approach, Pollock and Singh in Ref. \cite{Pollock} study the
thermodynamics of the de Sitter spacetime and quasi-de Sitter
spacetime. One of the quasi-de Sitter models studied presents
phantom-like behavior and the entropy associated to it is negative
defined. In Ref. \cite{phantom1}, an extension of these results is
found for a more general phantom-type spacetime leading to
negative entropies as well. In Ref. \cite{gonzalez-diaz}, a
different approach is given and positive entropies are found for
the phantom fluid by assuming that its temperature is negative
defined. Babichev et al. demonstrate in Ref. \cite{Babichev} that
the mass of a black hole decreases due to accretion of phantom
energy. Making use of this result, in Ref. \cite{phantom2} the
second law of thermodynamics ($dS_{Total}/dt> 0$) is studied in a
model of universe dominated by a phantom fluid in the presence of
black holes (the density of black holes being much smaller than
the density of phantom dark energy), concluding that it is not
preserved inside the event horizon.

Another approach applies local thermodynamics \cite{Bespro}. In
this paper we use it to study a Bose-Einstein gas of
self-interacting particles that produces acceleration, applying
statistical mechanics. Thus, our model has a detailed
microscopical description. Bose-Einstein condensates have been
widely studied in cosmology. The better candidates for cold dark
matter (axions, WIMPs, scalar fields) can be found in the Universe
in the form of Bose-Einstein condensates \cite{bec-cdm}. Also the
quantum field equations describing the Bose-Einstein condensate
allow to test experimentally some of the predictions of
semiclassical quantum gravity models of general relativity
\cite{bec-analog}. Here  we demonstrate how accelerated expansion
can be obtained by considering that the universe contains a
Bose-Einstein gas of self-interacting particles. Specifically, the
interaction between the particles of the gas gives accelerated
solutions (with $\rho+ 3p<0$, and even phantom-like solutions with
$\rho+p<0$). It is well known in the literature that the presence
of a self-interaction between elemental particles can originate an
accelerated expansion of the universe \cite{zimdahl0}. In
addition, Bose-Einstein gas is a feasible option to describe both
dark matter and dark energy, in contrast to radiation or ordinary
non-relativistic matter. We also demonstrate how this model can be
applied to describe the early stage of accelerated expansion,
solving the horizon problem and respecting the nucleosynthesis
scenario. In a future paper, we will demonstrate how present day
accelerated expansion can be achieved in our model.

In Sec. II, we use expansions within statistical mechanics to
obtain the equations of state of this gas. In Sec. III, we study
the dynamics of a Friedman-Robertson-Walker (FRW) universe that
contains the interacting Bose gas, making use of the results of
the previous section. We obtain the parameter conditions leading
to accelerated non-phantom and phantom solutions. In particular,
we study the dynamics of a FRW containing the gas, and then with
radiation. Then, we show how the horizon problem is solved for
solutions with $\rho+p<0$. We also investigate how the model
matches the standard cosmology, respecting the nucleosynthesis
scenario, and producing adequate fluctuations. Finally, we  study
the stability of the model. In Sec. IV, we summarize our findings.

We use units with $\hbar=k_B=c=1$.

\section{Short-range interactive Bose gas }

The thermodynamic potential $\Omega$  of a Bose gas, described by
single-particle quantum states with energy levels $\epsilon_k$
associated to momentum $k$ \cite{Landau}, is given by

\begin{eqnarray}
\label {thermopot} \Omega=T \sum_k \log(1-e^{(\mu-\epsilon_k)/T}),
\end{eqnarray}
where $T$ is the temperature, and $\mu$ the chemical potential. It
leads to the average occupation number
\begin{eqnarray}
\label{occupation} \bar n_k=  \frac{1}{e^{(\epsilon_k-\mu)/T}-1 },
\end{eqnarray}
where $\epsilon_k$ contains the kinetic energy of a particle
with momentum $k$ and mass $m$;
considering phase space, the number of particles associated to a
given state is
\begin{eqnarray}\label{numberpar}
d N_k=  g \frac{V}{(2  \pi)^3} \bar n_k d^3 k,
\end{eqnarray}
where for spin-zero particles the degeneracy factor is $g=1$ and
$V$ is the volume. The total number of particles is
 \begin{eqnarray}\label{totalnumberpar} N=\int d N_k .\end{eqnarray}

Consistency of Eq. (\ref{totalnumberpar}) requires  macroscopic
occupation of the lowest state at low enough temperature $T$. In
the free case, this starts at the critical temperature $T_c$ such
that

\begin{eqnarray}\label {Tcrit}
N/V=\frac{g V ( T_c m)^{3/2}}{\sqrt{2} \pi^2}  \int_0^\infty
dz\frac{z^{1/2}}{e^{z}-1},
\end{eqnarray}
or equivalently
\begin{eqnarray}\label {TcritExpli}T_c = 3.31 \frac{1}{g
^{2/3}m}\left(\frac{N}{V}\right)^{2/3}.\end{eqnarray} The chemical
potential remains constant and zero for $T<T_c$; the latter
condition is required for the condensate phase to be present. In
addition, a non-relativistic description is assured if $m\gg T_c$,
or equivalently, \begin{eqnarray}\label {masscondition} m\gg
\left(\frac{N}{V}\right)^{1/3}.\end{eqnarray}.

A short-range two-particle interaction $V(x)$ modifies
free-particle behavior. We consider  temperatures below the
critical temperature, $T \leq T_c$. Near $T_c$, and $T< T_c$, the
single-particle energy is given (to first-order) by Ref. \cite{
Walecka}

\begin{eqnarray}\label {singleparticle}
\epsilon_{k}=\frac{k^2}{2 m}+ 2 v_0 ^\prime  n_{\epsilon>0}  ,
\end{eqnarray}
where  the second term is the potential energy of $N_{\epsilon>0}$
non-condensate particles associated with $V(x)$,
$n_{\epsilon>0}=N_{\epsilon>0}/V,$ and

\begin{eqnarray}\label {shortrange}
v_0^\prime=\int d^3x V(x).
\end{eqnarray}
For  Eq. (\ref{singleparticle}), the chemical potential is

\begin{eqnarray}\label {chemicalpotenticalhighT}
\mu= 2 n_c v_0 ^\prime ,\end{eqnarray} instead of the $\mu=0$
value
 for a free gas with $T<T_c$. The value of $T_c$ in Eq.
(\ref{Tcrit}) and the constancy of $\mu$ remain  valid also in
this case.

A similar  constant potential term is approximated near total
condensation $T\sim 0$ \cite{Landau}  with $n_c=N_c/V$, $N_c$ the
condensate particle number,  for which
\begin{eqnarray}\label {singleparticlelowT}
\epsilon_{k}=\frac{k^2}{2 m}+  v_0^\prime n_c ,
\end{eqnarray}
and
\begin{eqnarray}\label {chemicalpotenticallowT}
\mu=   n_c v_0^\prime  .
\end{eqnarray}
The presence of a factor 2 within Eq.
(\ref{chemicalpotenticalhighT}) expresses the exchange effect,
which accounts for interaction of different states, while these
corrections are absent to this order for the condensate. We use
the limits in Eq. (\ref{singleparticle}) for $T$ near $T_c$  and
Eq. (\ref{singleparticlelowT}) for $T$ near 0 to construct an
interpolated potential term. This takes into account the potential
particle exchange between the condensate and non-condensate
components, so that an appropriate average is  $v_0 = v_0^
\prime[( N_c^ 2+2 (N-N_c)^2+2 N_c(N-N_c)]  / N^2$,
($N=N_c+N_{\epsilon>0}$).

One way to derive the energy corresponding to $\Omega$ is using
the thermodynamic identity

\begin{eqnarray}\label {thermoidentity}
E=-\mu \frac{\partial\Omega}{\partial\mu}-T
\frac{\partial\Omega}{\partial T}+\Omega.
\end{eqnarray}
Another way to obtain it in the interactive case is to substitute
the single-particle distribution  into the Bose distribution in
Eq. (\ref{numberpar}).  Then,

\begin{eqnarray}\label
{energymodelsecond} E&=&\frac{g V  m^{3/2}}{\sqrt{2} \pi^2} \int
d\epsilon
\epsilon^{3/2}\frac{1}{e^{(\epsilon_k-\mu)/T}-1} +\frac{1}{2V} \sum_{i\neq j}^{N} v_0  \\
&\simeq&\frac{g V  m^{3/2}}{\sqrt{2} \pi^2} T^{3/2} \int_0^\infty
dz \frac{z^{3/2}}{e^{z}-1} + \frac{v_0}{2 V}N^2,
\end{eqnarray}
where the second term in $E$ sums over pairs of interactions
averaged over volume from  Eq. (\ref{shortrange}), and Eq.
(\ref{energymodelsecond}) takes the thermodynamic limit.

The pressure can be obtained from the definition $p=-\left(
\partial E /\partial V \right)_{N,S}$ and is

\be p=\frac{2}{3}\frac{E}{V} +
\frac{1}{6}\upsilon_0\frac{N^2}{V^2}.\ee

We obtain the entropy of the gas $S$ in terms of the energy $E$,
number of particles $N$ and volume $V$ from the well-known
thermodynamic relation

\be TS=E+pV-N\mu,\label{termo1}\ee where the temperature is
$T=(\partial E/\partial S)_{V,N}$. In our case, Eq. (\ref{termo1})
reads

\begin{equation} \left(\frac{\partial E}{\partial
S}\right)_{V,N}S=\frac{5}{3}\left( E-\frac{1}{2}v_0\frac{N^2}{V}
\right), \end{equation}

which can be integrated to

\be S=C V^{\frac{2}{5}}\left( E-\frac{1}{2}v_0\frac{N^2}{V}
\right)^{\frac{3}{5}} \label{entropy}, \ee where
$C=\frac{5}{3}\left( 128 g \right)^{2/5} m^{3/5}$ is set from the
interactive-gas entropy dependence on the kinetic energy, which is
similar to that of the non-interactive gas entropy in Ref.
\cite{Landau}.

\section{Cosmology of the interacting Bose gas}

We consider a flat Friedmann-Robertson-Walker (FRW) universe with
line element
\[
ds^{2}=-dt^{2}+a(t)^{2}\left[ dr^{2}+r^{2}d\Omega_{\theta}
^{2}\right],
\]
where $a(t)$ is the scale factor, $t$, $r$ and $\Omega_{\theta}$
are the time, the radius and solid-angle comoving coordinates of
the metric, respectively. The Einstein equations \cite{Kolb} read

\be H^2 = \left( \frac{\dot{a} }{a}\right)^2 = \frac{8 \pi}{3
m_{P}^2}\rho, \label{eqH} \ee

\be \dot{\rho} + 3H(\rho+p) = 0, \label{evdens}\ee where $m_{P}$
is the Planck mass, $\rho$ and $p$ are the total energy density
and total pressure of the fluid conforming the universe,
respectively. For a universe containing $n$ fluids, the total
energy density and pressure are $\rho=\sum_{j=0}^n \rho_j$ and
$p=\sum_{j=0}^n p_j$, where $\rho_j$ and $p_j$ are the energy
density and pressure of each component $j$. If there is no
interaction between the different components, we can generalize
Eq. (\ref{evdens}) to a set of $n$ independent equations

\be \dot{\rho_j} + 3H(\rho_j+p_j) = 0. \label{evdensi}\ee

Combining Eqs. (\ref{eqH}) and (\ref{evdens}), we find the
acceleration parameter

\be \frac{\ddot{a}}{a} = -\frac{4 \pi}{3 m_P^2} ( \rho + 3 p ).
\label{acc}\ee One concludes from this equation that the universe
is undergoing an accelerated expansion if the fluid violates the
strong energy condition $\rho + 3 p > 0$. Additionally, we can
consider the dominant energy condition $\rho + p> 0$ that assures
that the speed of sound in the fluid be lower than $c$.

\subsection{Friedmann-Robertson-Walker metric containing the interacting Bose gas}

The pressure of the gas, from Eq. (\ref{entropy}), is

\be p=\frac{2}{3}\rho + \frac{1}{6}\upsilon_0\frac{N^2}{V^2}.\ee
In an expanding FRW universe the volume changes with time as
$V=V_i \left(a(t)/a_i\right)^3$, where $V_i$ and $a_i$ denote the
volume and scale factor at an arbitrary reference instant $t_i$,
respectively. Assuming a FRW universe model that expands
adiabatically, the above equation reads

\be p=\frac{2}{3}\rho + \frac{1}{6}\upsilon_0 n_i^2
\left(\frac{a_i}{a(t)}\right)^{6},\label{pc}\ee where $n_i$ is the
number density of particles (i.e., $n_i=N/V_i$, where $N$ is
defined in Eq. (\ref{totalnumberpar})) at $t_i$. Using the
equation of state (\ref{pc}) in  Eq. (\ref{evdens}), we obtain the
evolution of the energy density

\be
\rho=\rho_{ci}\left(\frac{a_i}{a(t)}\right)^{5}+\frac{1}{2}\upsilon_0
n_i^2 \left(\frac{a_i}{a(t)}\right)^{6}, \label{ed0}\ee where
$\rho_{ci}$ denotes the energy density of the condensate at the
reference instant $t_i$.

The Bose-gas equality  $T=(5/3) E_c/S$  (Ref. \cite{Landau}) for
the condensate energy contribution $E_c=\rho_c V$ confirms that
once the condition $T< T_c$ is  satisfied at some point, it is
always satisfied, as both  $T$  (Eq. (\ref{ed0})) and $T_c$ (Eq.
(\ref{TcritExpli})) decrease as $a^{-2}$. This also means the
non-relativistic condition in Eq. (\ref{masscondition}) can be
consistently maintained throughout.

From Eq.(\ref{acc}), the universe acceleration containing the gas
is

\be \frac{\ddot{a}}{a} = -\frac{4 \pi}{3 m_P^2} \left[ 3
\rho_{ci}\left(\frac{a_i}{a(t)}\right)^{5}+2 \upsilon_0 n_i^2
\left(\frac{a_i}{a(t)}\right)^{6} \right]. \label{accA}\ee It is
possible to obtain accelerated solutions to the Einstein equations
if the interaction term is negative ($\upsilon_0<0$), i.e., if the
interaction is attractive. From this point on, we assume $v_0<0$
\footnote{In this case, the second term in the right hand side of
Eq. (\ref{pc}) can be interpreted as an effective viscous
pressure. The dynamics of a universe containing a viscous fluid
has been widely studied and its presence allows  the expansion to
accelerate under certain conditions \cite{Mendez},
\cite{zimdahl}.}. When the acceleration given by Eq. (\ref{accA})
is positive at $a=a_i$ (i.e., $-2 \upsilon_0 n_i^2> 3 \rho_{ci}$),
the model goes through a transitory stage of accelerated expansion
that ends when the interaction term (which scales as $a^{-6}$)
becomes negligible as compared to the condensate term (which
scales as $a^{-5}$). We evaluate the scale factor for which the
accelerated stage ends (i.e., $\ddot{a}(a/a_i)=0$). This is at \be
\frac{a}{a_i}=-\frac{2\,\upsilon_0 n_i^2}{3\,\rho_{ci}}.
\label{a/a_iNoRad} \ee General Relativity (GR) requires that the
total energy density be positive, i.e., $-\upsilon_0
n_i^2<2\rho_{ci}$. Using this condition in Eq. (\ref{a/a_iNoRad}),
we conclude that the accelerated expansion stage ends at a scale
factor $a/a_i$ always lower than $4/3$. The transitory
acceleration stage predicted by this model is hence more suitable
to describe the acceleration era of the early Universe than the
present accelerated era. To describe the present era of expansion
with this model, the interaction term should evolve with the scale
factor in a different way to become the dominant term of the
expansion later in time. This will occur if the interaction
parameter $v_0$ evolves with time, or if the number density of the
gas particles $n$ is not conserved with the expansion.

We now explore the conditions under which the transitory
accelerated stage predicted by this model can describe the
acceleration of the early Universe, and in particular, solve the
horizon problem.

\subsection{Friedmann-Robertson-Walker metric containing the interacting Bose
gas and radiation}

To address this application of the accelerated stage, we add to
the model a radiation term with density
$\rho_r=\rho_{ri}\left(a_i/a(t)\right)^{4}$ and pressure
$p_r=\rho_r/3$. This term is introduced to fit the standard
cosmology, as indeed it eventually dominates the expansion,
allowing for nucleosynthesis to happen.

The total energy density and pressure of the universe are

\be
\rho=\rho_{ri}\left(\frac{a_i}{a(t)}\right)^{4}+\rho_{ci}
\left(\frac{a_i}{a(t)}\right)^{5}+\frac{1}{2}\upsilon_0
n_i^2 \left(\frac{a_i}{a(t)}\right)^{6}, \label{ed0rad}\ee

\be p=\frac{1}{3}\rho_{ri}
\left(\frac{a_i}{a(t)}\right)^{4}+\frac{2}{3}\rho_{ci}
\left(\frac{a_i}{a(t)}\right)^{5}+ \frac{1}{2}\upsilon_0 n_i^2
\left(\frac{a_i}{a(t)}\right)^{6}.\label{pcrad}\ee The free
parameters in this model are the density of the radiation
$\rho_{ri}$, the energy density of the condensate $\rho_{ci}$, and
the interaction term $\upsilon_0 n_i^2$. The parameters are
constrained by the condition $-\upsilon_0
n_i^2<2(\rho_{ri}+\rho_{ci})$, as the total energy density
entering the Einstein equations is required to be positive, i.e.,
$\rho(a_i)>0$. The reference term $a_i$ can be interpreted as the
scale factor at which our model starts to be valid.

The acceleration reads

\be \frac{\ddot{a}}{a} = -\frac{4 \pi}{3 m_P^2} \left[ 2
\rho_{ri}\left(\frac{a_i}{a(t)}\right)^{4}+3
\rho_{ci}\left(\frac{a_i}{a(t)}\right)^{5}+2 \upsilon_0 n_i^2
\left(\frac{a_i}{a(t)}\right)^{6} \right]. \label{accB} \ee
Similarly to the model in subsection 3.1, it is possible to obtain
accelerated solutions to the Einstein equations if $\upsilon_0<0$.
The acceleration eventually becomes negative, as the interaction
term scales as $(a_i/a)^6$, while the contributions of the
radiation and the condensate scale as $(a_i/a)^4$ and $(a_i/a)^5$,
respectively. When the latter terms dominate, the expansion
becomes decelerated.

The left and right panels of Fig. \ref{fig1} show the evolution of
the total density and acceleration, respectively, for three
possible types of solutions, which are characterized by the
acceleration at $a_i$. These are:

\renewcommand{\labelenumi}{\roman{enumi}.}
\begin{enumerate}
\item{Decelerated solutions}

The acceleration given by Eq. (\ref{accB}) is negative (or zero)
at the scale factor $a=a_i$ when $\rho(a_i)+3p(a_i)>0$, i.e.,
$-\upsilon_0 n_i^2 \leq (3/2)\rho_{ci}+\rho_{ri}$. Then, the
universe maintains a decelerated expansion from $a=a_i$ on.

\item{Non-phantom accelerated solutions}

These solutions are obtained when the free parameters fulfil the
conditions $(3/2)\rho_{ci}+\rho_{ri}< -\upsilon_0
n_i^2<(5/3)\rho_{ci}+(4/3)\rho_{ri}$, where the lower limit is
given by the positiveness of the acceleration in Eq. (\ref{accB})
at the scale factor $a=a_i$, and the upper limit by the condition
$\rho+p>0$.

\begin{figure}[tbp]
\includegraphics*[scale=0.35]{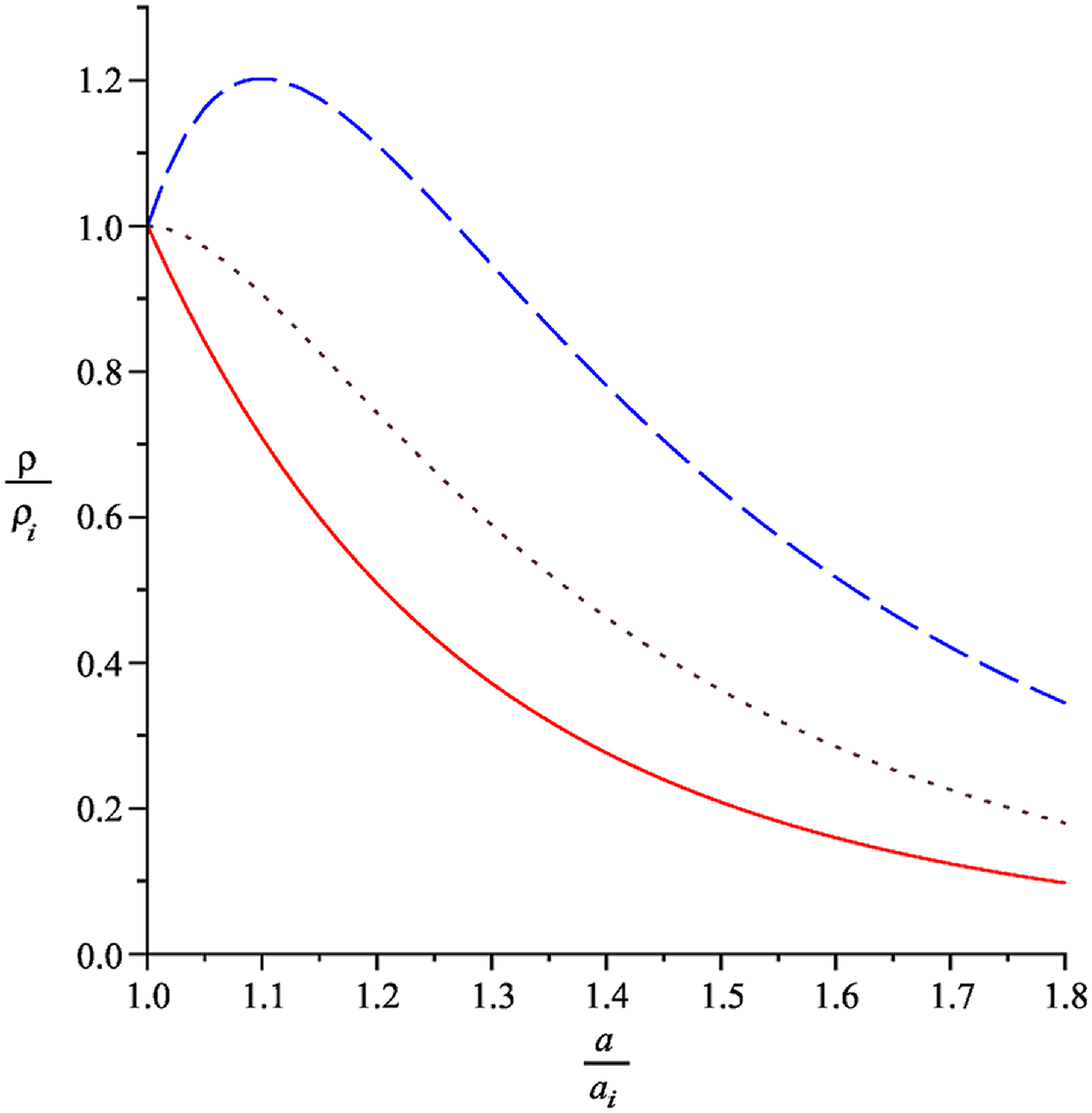}
\includegraphics*[scale=0.35]{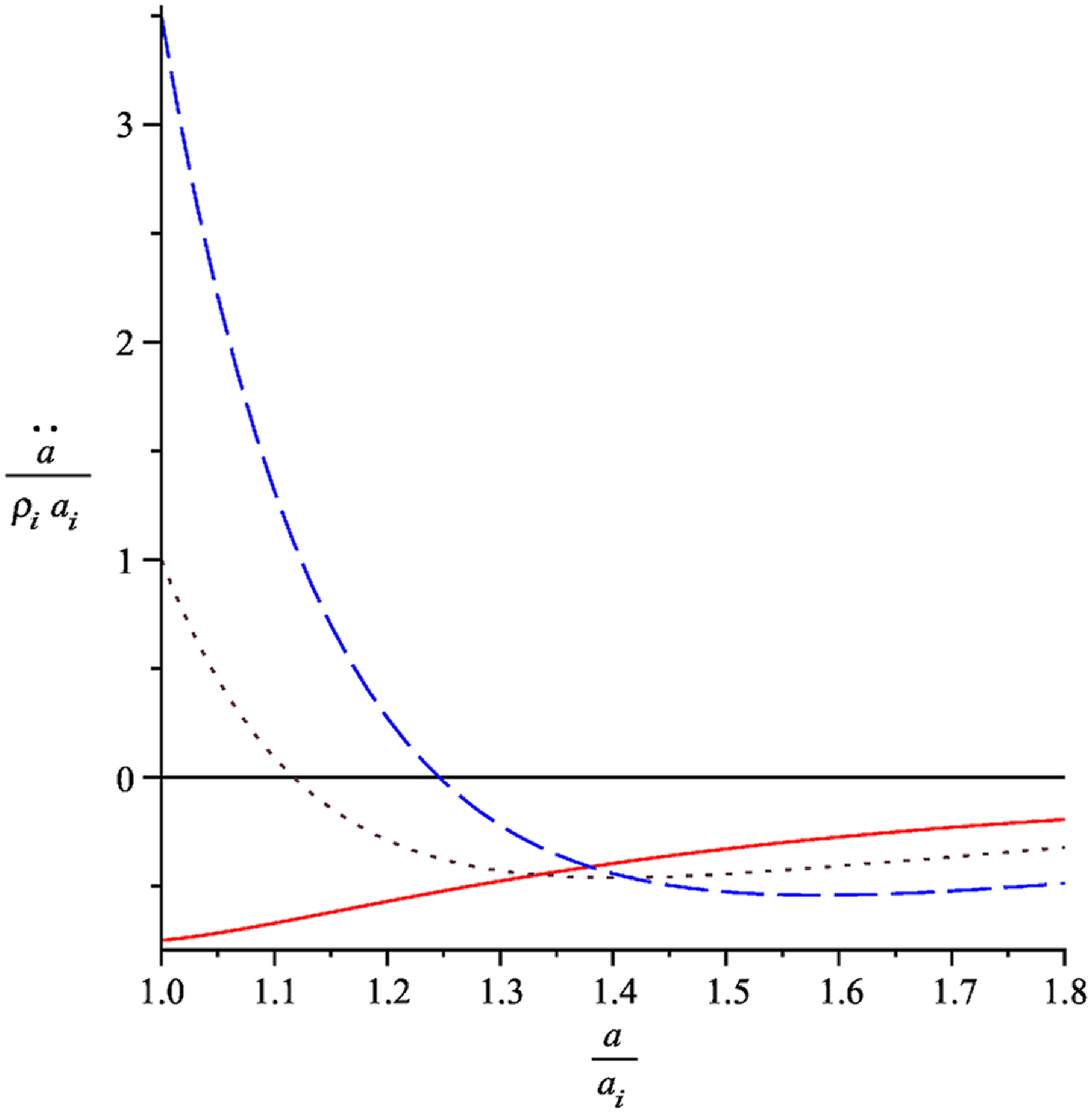} \caption{Left panel: Energy-density
evolution from the reference scale factor $a_i$ on, for different
choices of the free parameters. Right panel: Acceleration evolution
with the scale factor from the reference scale $a_i$ on, for different
choices of the free parameters. Every line is normalized to the value
of the density at $a=a_i$,
$\rho_i=\rho_{ri}+\rho_{ci}+(\upsilon_0 n_i^2)/2$. The three
different regimes are shown: i. Decelerated universe with
parameters $\rho_{ri}=0.5\rho_i$, $\rho_{ci}=1.5\rho_i$ and
$(\upsilon_0 n_i^2)/2=-\rho_i$ (solid line); ii. Accelerated
solution with parameters $\rho_{ri}=0.5\rho_i$,
$\rho_{ci}=5\rho_i$ and $(\upsilon_0 n_i^2)/2=-4.5\rho_i$ (dotted
line); iii. Super-exponential accelerated solution with parameters
$\rho_{ri}=3\rho_i$, $\rho_{ci}=5\rho_i$ and
$(\upsilon_0 n_i^2)/2=-7\rho_i$ (dashed line). We assume for this
and the next plots, $8 \pi/(3m_P^2)=1$.} \label{fig1}
\end{figure}

\item{Phantom-accelerated solutions}

The acceleration given by Eq. (\ref{accB}) at $a_i$ is positive, and
the condition $\rho+p>0$ is violated at the scale $a=a_i$, if
$-\upsilon_0n_i^2>(5/3)\rho_{ci}+(4/3)\rho_{ri}$. In this case, the
total density grows with the scale factor as
$d\rho/da=-3(\rho+p)/a>0$. First, the universe goes through a
super-exponential acceleration phase similar to that in the
phantom-energy models. When the interaction term is diluted
for large values of $a$, the condition $\rho+p<0$ is no longer
valid and the energy density becomes a decreasing function of
$a$. Consequently, the energy density presents a maximum
$\rho_{max}$. The scale factor $a_{*}$ for which
$\rho(a_{*})=\rho_{max}$ is \be
\frac{a_{*}}{a_i}=\frac{-6\upsilon_0
n_i^2}{5\rho_{ci}+\sqrt{25\rho_{ci}^2-48\upsilon_0
n_i^2\rho_{ri}}}.\label{a*/a0} \ee The plausibility and
stability of phantom models has been widely studied \cite{carrol}.
The big-rip problem (infinite expansion of the universe in a finite-time span)
in some phantom-energy models that explain the current
Universe acceleration \cite{caldwell} is not present in our model
because the phantom-accelerated stage is transitory and ends before
the scale factor diverges.
\end{enumerate}
The scale factor $a$ at which the accelerated expansion stage ends
in this scenario is always lower than $\sqrt{2}a_i$ (see the
Appendix A for details).

\subsection{Solving the horizon problem through the super-exponential expansion regime}

Next, after briefly reviewing the horizon problem, we obtain the
parameter conditions which solve horizon problem, and are
compatible with the nucleosynthesis scenario.  Also, the spectrum
and evolution of the initial density perturbations is addressed
later.

Cosmic microwave background (CMB) photons from different regions
of the sky are in thermal equilibrium at temperatures with
relative fluctuations up to $10^{-5}$. The most natural
explanation for this phenomenon is that photons from different
regions had been thermalized by being in causal contact
previously. However, the dynamics of a FRW universe containing
only radiation and/or non-relativistic matter (dust) cannot
describe such a situation as there was no time for those regions
to interact before the photons were emitted. In other words, the
distance light could travel from the instant the photons were
emitted $t_{e}$ until the photons were released at the instant
$t_{dec}$ is much smaller than the horizon distance today ($t_0$),
i.e., \be \label{Horizoncondition}
a_0\int_{t_{e}}^{t_{dec}}\frac{dt'}{a(t')}=a_0
\int^{a(t_{dec})}_{a(t_{e})}\frac{da'}{{a'}^2 H(a')}\ll a_0
\int^{t_0}_{t_{dec}}\frac{dt'}{a(t')}, \ee where
$a_0$ represents scale factor today. This problem is known in the
literature as the horizon problem \cite{Kolb}, \cite{Liddle}.

An alternative description of this problem considers the Hubble
radius $1/H$, where $H$ is defined by Eq. (\ref{eqH}). The Hubble
radius is closely related to the horizon \cite{dodelson}. A point
in space is no longer in causal contact with an observer when
its distance to it, $\lambda$, becomes larger than $1/H$. In
radiation and dust dominated universes, $1/H$ increases with a
slope with respect to $a$ larger than one. This implies that a
point A in space for which $\lambda_A=1/H_0$ (in causal
contact with the observer for the first time today $t_0$) was
never before in causal contact with the observer (i.e.,
$\lambda_A>1/H(a)$ for any $a<a_0$).

\begin{figure}[tbp]
\includegraphics*[scale=0.5]{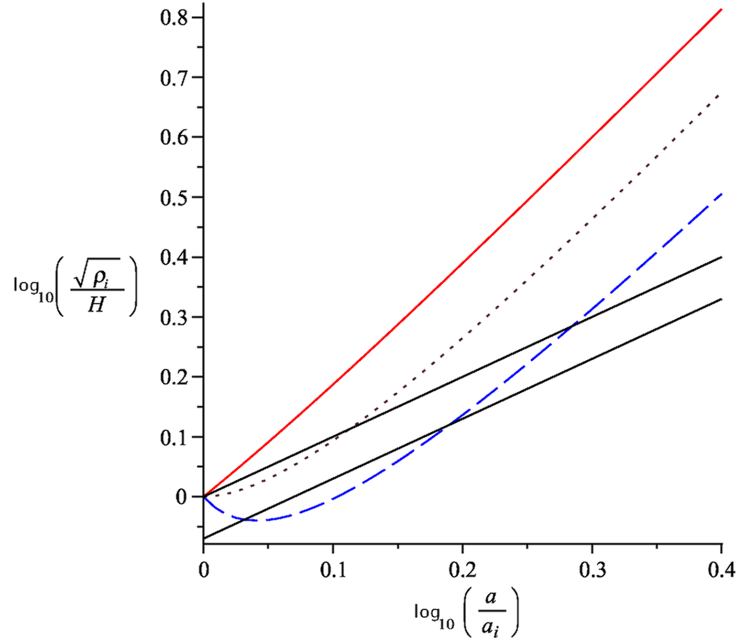}
\caption{Two different length scales that evolve with scale factor
as $\lambda=\lambda_i(a(t)/a_i)$ (black solid lines). The Hubble
radius, $1/H$, is plotted for different choices of the parameters:
1. Decelerated universe with parameters $\rho_{ri}=0.5\rho_i$,
$\rho_{ci}=1.5\rho_i$ and $(\upsilon_0 n_i^2)/2=-\rho_i$ (solid
line); 2. Accelerated solution with parameters
$\rho_{ri}=0.5\rho_i$, $\rho_{ci}=5\rho_i$ and $(\upsilon_0
n_i^2)/2=-4.5\rho_i$ (dotted line); 3. Super-exponential
accelerated solution with parameters $\rho_{ri}=3\rho_i$,
$\rho_{ci}=5\rho_i$ and $(\upsilon_0 n_i^2)/2=-7\rho_i$ (dashed
line). For this plot, $\rho_i=\rho_{ri}+\rho_{ci}+(\upsilon_0
n_i^2)/2$.
} \label{fig2}
\end{figure}
In both the accelerated and phantom-accelerated cases of the FRW
model containing radiation and the Bose gas exposed in the
previous subsection, there is an initial period for which
different length scales $\lambda$ are in casual contact, leaving
the Hubble radius and reentering it later in time, when the universe
is no longer accelerated. This process is illustrated in Fig.
\ref{fig2}. For the accelerated region, $d(1/H)/(da)$ is lower
than 1 but positive until the expansion becomes decelerated. For
the super-exponential accelerated region, $d(1/H)/(da)$ is
negative until the expansion becomes decelerated. This allows for
a large range of length scales to leave the causal contact region
and to reenter it later. Finally, we also note that for the
decelerated region, this phenomenon is not manifested, as
$d(1/H)/(da)$ is always larger than one (the particle horizon
tends to zero when $a$ decreases, as for the radiation or the
non-relativistic matter FRW models).

\begin{figure}[tbp]
\includegraphics*[scale=0.5]{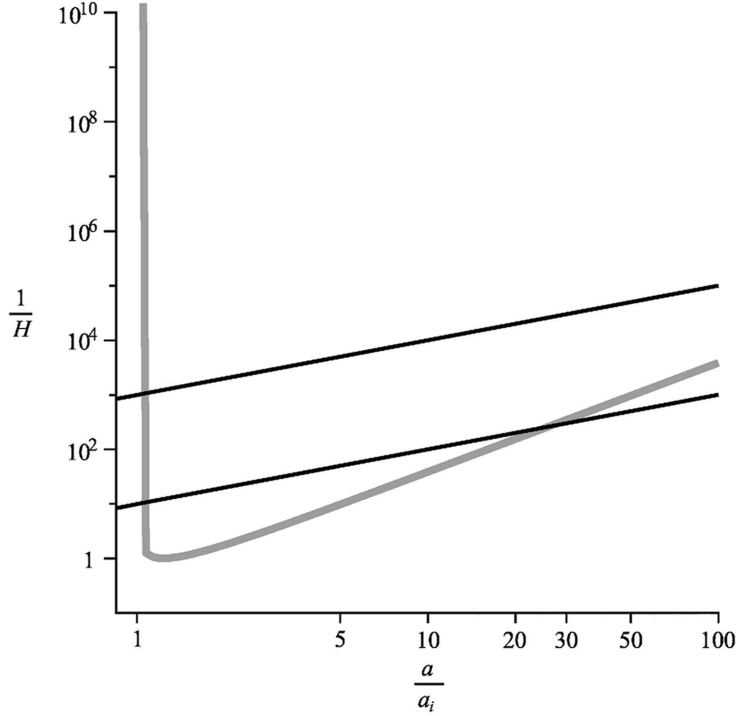}
\caption{$1/H$ as a function of $a/a_i$ in log-log plot for a set
of parameters that allows $1/H_i=10^{30}/H(a_{*})$. The solid
lines, represent two different length scales that evolve with
scale factor as $\lambda=\lambda_i(a(t)/a_i)$. } \label{fig3}
\end{figure}
It is possible to solve the horizon problem if we consider
phantom-accelerated values for the free parameters of our model;
specifically, we choose the scale factor $a_i$ to be small enough
so that the Hubble radius is big enough. In this scenario,
choosing $1/H_i\gg1/H(a_{*})$ ($a_{*}/a_i$ defined by Eq.
(\ref{a*/a0})), the distance that light can travel before the
photons were released can be large enough to solve the horizon
problem because

\be \int^{a_{*}}_{a_{i}}\frac{da'}{{a'}^2H(a')} \gg
\int^{t_0}_{t_{dec}}\frac{dt'}{a(t')}. \ee Then, we ensure that  there
be wide range of scales that leave the region of causal contact.
Fig. \ref{fig3} illustrates an example of a phantom regime for
which $1/H_i=10^{30}/H(a_{*})$. This implies that the energy density
increases from $\rho_i$ until it reaches its maximum
$\rho_{max} \approx 10^{60} \rho_i$.

\begin{figure}[tbp]
\includegraphics*[scale=0.65]{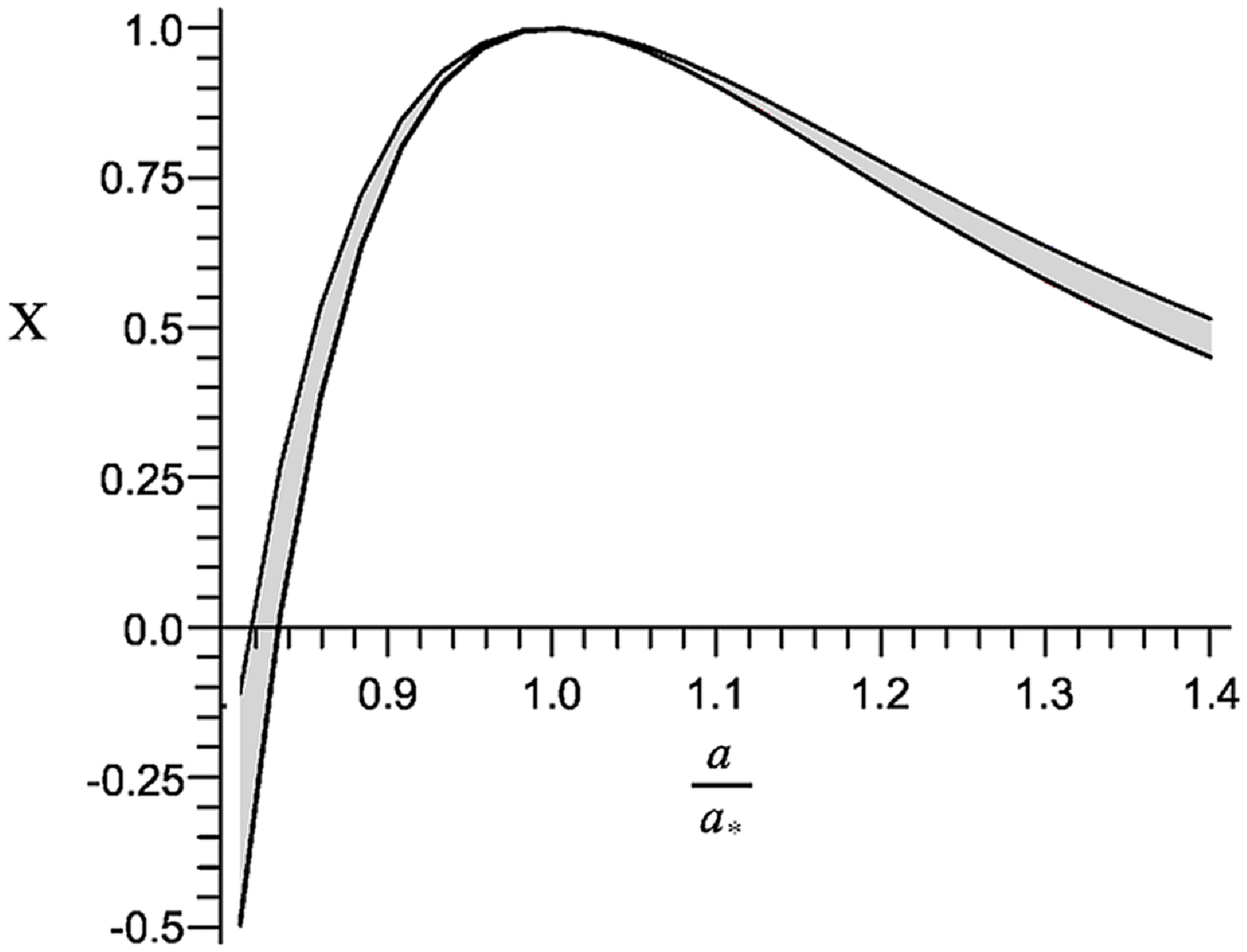}
\caption{The energy density as a function of the scale factor. The
shaded zone represents the shapes of the energy density depending
on the value free parameter $X^{*}_{v}$. The contours of the
shaded zone are the limit values of the free parameter: when the
condensate term is null ($X^{*}_{v}=2$); and when the radiation
term is null ($X^{*}_{v}=5$).} \label{fig4}
\end{figure}

We evaluate the range of the free parameters that solve the
horizon problem. It is plausible to assume that $\rho_{max}$
cannot be larger than the Planck scale, $m^4_{P}$ (GR is valid for
$\rho<m_P^4$). We first define the new set of parameters

\ben X^{*}_{r}=\frac{\rho_{ri}}{
\rho_{max}}\left(\frac{a_i}{a_{*}}\right)^{4},\,
X^{*}_{c}=\frac{\rho_{ci}}{
\rho_{max}}\left(\frac{a_i}{a_{*}}\right)^{5},\\
\qquad \, \, X^{*}_{v}=-\frac{1}{2}\frac{\upsilon_0 n_i^2}{
\rho_{max}}\left(\frac{a_i}{a_{*}}\right)^{6}. \label{rescale1}
\een
$X^{*}_{v}$ has been defined to be positive when the
interaction is attractive. These parameters are the contribution
of each energy component at $a_*$ to $\rho_{max}$. We define also
\be X(a)=\frac{\rho(a)}{\rho_{max}}\ee as the total energy density
reescaled to $\rho_{max}$. Its evolution in terms of the scale
factor can be expressed as

\be X(a)=X^{*}_{r}\left(\frac{a_{*}}{a}\right)^{4}
+X^{*}_{c}\left(\frac{a_{*}}{a}\right)^{5}-X^{*}_v
\left(\frac{a_{*}}{a}\right)^{6}.\ee The rescaled parameters are
related because $X(a_*)=1$, that is,

\be X^{*}_{r}+X^{*}_{c}-X^{*}_{v} =1. \label{cond1} \ee In
addition, the derivative of the total density respect to $a$
evaluated at $a_{*}$ is equal to zero. This condition reads

\be -4X^{*}_{r}-5X^{*}_{c}+6X^{*}_{v}=0. \label{cond2} \ee
Combining conditions (\ref{cond1}) and (\ref{cond2}), we relate
the parameters in the form
 \be X^{*}_{r}=5-X^{*}_{v},\qquad
X^{*}_{c}=2X^{*}_{v}-4,\label{parameters}\ee and, then, the only
free parameter to consider is the interaction $X^{*}_{v}$. Given
Eq. (\ref{parameters}) and because $X^{*}_{r}>0$, $X^{*}_{v}<5$.
Additionally, $X^{*}_{c}>0$, which implies that $X^{*}_{v}>2$. The
range of parameters that can solve the horizon problem in our
model are then given by Eq. (\ref{parameters}) with
$2<X^{*}_{v}<5$. Fig. \ref{fig4} shows the energy density as a
function of the scale factor, for this range of parameters. The
energy density reaches the null value at the scale \be
\frac{a_{**}}{a_{*}}=\frac{X^{*}_{v}}{X^{*}_{v}-2+\sqrt{X^{*}_{v}+4}},
\ee and the horizon problem is solved, and Eq.
\ref{Horizoncondition} is satisfied, because the energy density
increases from $\rho(a)\approx 0$ at a scale factor $a \gtrsim
a_{**}$. Note that if we consider $a_{**}$ as the initial scale,
$1/H(a_{**})$ diverges, and so does the horizon. This divergence
is also found in the inflationary de-Sitter models, for which the
horizon diverges when $a(t_e) \rightarrow 0$.

\subsection{Fluctuations}

One of the main predictions of inflation is the presence of
scale-invariant fluctuations which form the seed of later large-scale
structure formation. In the conventional  description
for a nearly de Sitter  universe, the constancy of the fluctuations
amplitude  is determined  by the constant horizon size (which in
turn is set by the Hawking temperature $T_H = H/(2 \pi)$); this
occurs at horizon scale $k=H a$, where $k$ is the
Fourier-transformed comoving coordinate.

We have shown that the universe can be initially causally
connected in our model; we shall also  argue   that the properties
of a Bose gas allow for a scale-invariant perturbations regime: we
show that thermal fluctuations originating in the non-condensate
contribution have the appropriate properties for  the standard
cosmology.

During  the phantom-accelerated regime,  the expansion is dominated
first by the interaction term in Eq. (\ref{ed0rad})  and the
non-condensate contribution; the latter then dominates as $a$ grows
(see Fig. 1). In general, the   mean square density fluctuation, for
given volume (and therefore, constant particle number or entropy),
is given by \cite{Callen}

\begin{eqnarray}\label {EnergyFluc}
\langle (E-\bar E)^2 \rangle = T^2 \left ( \frac{\partial
E}{\partial T}\right )_{N,V}=T^2 N c_v,
\end{eqnarray}
where  $c_v$ is the specific heat per particle (constant in our case).

While this calculation represents a useful benchmark for later
comparison, we actually need the contribution of the fluctuations at
given mode $k$. This may be obtained from the approximate
large-distance non-condensate  wave-function component $\delta \Phi_k$
\cite{LandauII}

\begin{eqnarray}\label {EnergyFlucsumak}
\delta \Omega = \frac{1}{2 m^2 }\rho_m V \sum_{\bf k_{phys}}  k_{phys}^2|\delta
\Phi_k |^2,
\end{eqnarray}
where $\rho_m$ is the non-condensate mass density, and we used the
physical $k_{phys}$; this term contributes to the kinetic term in
Eq.  (\ref{ed0rad}).
Then, one  obtains the  expresion for the statistical average
\begin{eqnarray}\label {EnergyFlucsumakTot}
 \langle|\delta \Phi_k |^2 \rangle =\frac{T m^2}{V  \rho_m k_{phys}^2}.
\end{eqnarray}
By considering the  phase-space factor $V d^3k_{phys}=  d^3k$, one
finds the  $k$ contribution  to the total fluctuation in Eq.
(\ref{EnergyFluc}) is proportional to $T$.
While $T$ is $a$-dependent, the ratio in the rms density contrast is
not, %cita ecuacion, libro
nor is $\zeta_{rms}=\delta\rho_{rms}/(\rho+p)$, as one can use the
relation $p_c=(2/3)\rho_c$ for the non-condensate part. We hence obtain
that the gauge-invariant quantity contribution $\zeta_{rms}$ is
constant beyond the horizon. This leads to gaussian scale-invariant
fluctuations, since their origin is thermodynamic, and they all have
the same $k$-power dependence for their amplitude.

Once the fluctuation  mode $k$ crosses outside the Hubble radius,
it evolves as a classical perturbation \cite{Kolb}; the
fluctuation evolution of the universe in our model becomes
equivalent to the standard universe. Thus, we find that it  is
possible to obtain scale-invariant fluctuations, assuming the
non-condensate fluctuations dominate, even as the horizon size
$1/H$ is not constant, but decreases with $a$ (or equivalently
with $t$).

\subsection{Nucleosynthesis}

We find the condition under which our model solves the horizon
problem consistently with standard cosmological requirements, and
specifically with nucleosynthesis \cite{Kolb}. Nucleosynthesis
requires that the Universe be dominated by radiation at times of
around $1 s$ in order for the light elements to be produced.

We consider the scale factor $a_{eq}$ at which the radiation and
Bose-gas densities are equal, i. e.,

\be X_{r}^{*}\left( \frac{a_{*}}{a_{eq}}\right)^4=X_{c}^{*}\left(
\frac{a_{*}}{a_{eq}}\right)^5-X_{i}^{*}\left(
\frac{a_{*}}{a_{eq}}\right)^6. \label{aeq} \ee From the scale
$a_{eq}$ on, the radiation term will dominate the expansion of the
universe. We evaluate the time span $t_{eq}$ which corresponds to
this scale factor. If $t_{eq}<1s$ then the nucleosynthesis
scenario will be preserved by the model. The time span $t_{eq}$ is
found by integration of Eq. (\ref{eqH}),
 \be
t_{eq}=\sqrt{\frac{3}{8 \pi}}\sqrt{\frac{m_P^2}{\rho_{max}}}
\int^{\frac{a_{eq}}{a_{*}}}_{\frac{a_{**}}{a_{*}}} \frac{d\left(
\frac{a}{a_{*}}\right)}{\frac{a}{a_{*}} \sqrt{X\left(
\frac{a}{a_{*}}\right)}}. \label{teq} \ee

If there is no radiation in our model, the parameters read
$X^{*}_{r}=0$, $X^{*}_{c}=6$, $X^{*}_{v}=5$. In this particular
case, $t_{eq}$ tends to infinity. These parameters, although they
solve the horizon problem, are in contradiction with the
nucleosynthesis scenario.

Depending on the value of $\rho_{max}$ it is possible to
numerically evaluate how much radiation density is initially
needed (i.e., how large $X^{*}_{r}$ must be) to allow for
$t_{eq}<1s$. This minimal radiation amount decreases its value as
$\rho_{max}$ increases, as can be seen from Eq. (\ref{teq}).

Assuming $\rho_{max}=m_P^4$ and that $X^{*}_{r}\approx 10^{-17}$,
$X^{*}_{c}\approx 6-2 \times 10^{-17}$, $X^{*}_{v}\approx
5-10^{-17}$, then Eq. (\ref{teq}) reads $t_{eq}\approx 10^{43}
m_P^{-1} \approx 0.1 s$. In this limit, nucleosynthesis will be
allowed for any $X^{*}_{r} \geq 10^{-17}$. If we consider that
$\rho_{max}$ is lower than $m_P^4$, the minimal amount of radiation
needed (evaluated from Eq. (\ref{teq}) with $t_{eq}=1s$) is larger
than $10^{-17}$.

Consideration of the Bose-gas particles' mass requires the
contribution $m n$ in the energy density. This term scales as
$a^{-3}$, and eventually becomes  the dominant contribution. To
ensure that this term does not dominate the expansion before
nucleosynthesis, we have to impose the condition $mn_i \ll
\rho_{ri}$. The latter condition, together with the
non-relativistic gas condition given by equation
(\ref{masscondition}), implies $\rho_{ci}\ll \rho_{ri}, (v_0
n_i^2)/2$. In terms of the rescaled parameters, this is $X^{*}_{c}
\approx 0$, $X^{*}_{v}\approx 2$ and $X^{*}_{r}\approx 3$.

We can conclude that nucleosynthesis does not bound the free
parameters of the model strongly. If the radiation term does not
differ from the other two terms for more than 10 orders of
magnitude, nucleosynthesis will occur. In such a scenario, our
model starts with an accelerated era that explains the horizon
problem, later evolves in a radiation dominated universe in which
light elements can be created and, eventually, the expansion is
dominated by non-relativistic matter. In other words, our model of
universe behaves as the standard cosmology, besides the initial
stage of expansion. As we mentioned before, late time
acceleration can be also obtained in our model if we consider that
the number of particles of the gas or the interaction term are not
constant. This problem will be addressed in a future paper.

\subsection{Stability of the interacting Bose gas in the cosmological frame}

A  confined Bose-gas condensate with an attractive interaction is
stable under certain conditions \cite{Dalfovo}, unlike the case of a
Bose condensate with an attractive interaction within an infinite
volume, which always collapses. When the gas' particles are confined
in a given volume (e.g., by a  harmonic trap), finite-size effects
due to the confining potential generate an additional kinetic-energy
contribution that compensates the interaction energy, avoiding the
collapse, as long as the confined  particle number   in the
potential is lower than a critical value $ N_{cr}$. This number can
be computed by equating the exterior potential plus kinetic energy
terms with the interaction term \cite{Dalfovo}, which translates to
\be N_{cr} \lesssim 4 \pi \frac{a_{ho}}{m|v_0|}\label{quantcond},\ee
where $a_{ho}$ is the physical trap characteristic length scale, and $m$
is the  particle mass.

We approach the collapse problem in our model applying dimensional
analysis. In the cosmological frame, we assign $a_{ho}$ to the
Hubble radius $1/H$, as the latter represents the region of space
in causal contact. In our case, $a_{ho}$ grows with time, as the
volume expands \cite{Kim}. Thus, if the Hubble volume $(1/H)^3$
contains less than $N_{cr}$ particles, the collapse is avoided. So
is the case for any smaller volume with a characteristic length
$a_{hol}$ lower than $1/H$, as such a volume will contain less
than $N_{cr}$ particles.

It is sufficient to apply the above condition for initial values of
the parameters in our  model, as the interaction energy term
(proportional to $n^2$) dilutes faster with the expansion than the
condensate term in Eq. (\ref{ed0rad}), and eventually becomes
negligible.  With the association of $N_{cr}$ to the ground-state
(or condensate)  particle density $n_c$, Eq. (\ref{quantcond}) reads

\be n_i^2|v_0| \lesssim  \frac{\rho_{ri}+\rho_{ci}}{
\frac{1}{2}+\frac{3m_P^2 m n_c}{2(4\pi)^2 n_i^2}  },
\label{colapsecond}\ee where  we used $ a_{ho} \sim
1/H_i=\{[8\pi/(3m_P^2)](\rho_{ri}+\rho_{ci}+v_0n_i^2/2)\}^{-1/2}$,
and  $n_i$ is the total initial particle density. Using the
non-relativistic condition for the mass in Eq. (\ref{masscondition}),
$m \gg n_i^{1/3}$, and having in mind that $m<m_P$, it is convenient
to explore two extreme cases

\renewcommand{\labelenumi}{\roman{enumi}.}
\begin{enumerate}

\item{$n_i \gg n_c$}

Only a small amount of the total particles are in the ground state.
As the temperature of the gas is
$T=T_c\left(\frac{N-N_c}{N}\right)^{2/3}$ \cite{Landau}, we have $T
\sim T_c$. If   $m$   is such that the second term in the
denominator in  Eq. (\ref{colapsecond}) satisfies \be \frac{3m_P^2 m
n_c}{2(4\pi)^2 n_i^2}\ll1, \ee then Eq. (\ref{colapsecond}) reads
$n_i^2|v_0|\lesssim2(\rho_{ri}+\rho_{ci})$, which is consistent with
the condition required for the total energy density entering the
Einstein equation to be positive. In this case, it is possible to
obtain phantom-accelerated, non-phantom accelerated and decelerated
solutions that do not  collapse.

\item{$n_i=n_c$}

All the particles of the gas are in the ground state (i.e., the
temperature of the gas is $T=0$,)  and consequently, \ben
\frac{3m_P^2 m n_c}{2(4\pi)^2
n_i^2}=\frac{3}{2(4\pi)^2}\frac{m_P^2 m}{n_i} \gg 1  \nonumber \\
\Rightarrow \rho_{ri}+\rho_{ci}\gg \frac{\rho_{ri}+\rho_{ci}}{
\frac{1}{2}+\frac{3m_P^2 m n_c}{2(4\pi)^2 n_i^2}}\gtrsim
n_i^2|v_0|. \een In this case
$n_i^2|v_0|\ll\rho_{ri}+\rho_{ci}$. The condition of $\rho>0$ is
compatible with the latter condition. But conditions for the
accelerated and phantom kind solutions in Section 3.2
($(3/2)\rho_{ci}+\rho_{ri}< -\upsilon_0
n_i^2<(5/3)\rho_{ci}+(4/3)\rho_{ri}$ and $-\upsilon_0
n_i^2>(5/3)\rho_{ci}+(4/3)\rho_{ri}$, respectively) are in
contradiction with the collapse condition in this case.
Consequently, models for which $n_i=n_c$ and that do not
present collapse are suitable to be used in cosmology (as
$\rho>0$) but lead always to a decelerated expansion of the
universe.

\end{enumerate}
The above  conditions are further relaxed in a cosmological
context, as an attractive interaction ultimately accelerates the
expansion, and temperature effects will contribute to counter the
attraction. To summarize this subsection, stable solutions of the
Bose-Einstein gas are possible, while accelerated ones favor the
case $n_i \gg n_c$.

\section{Conclusions}

A Bose-Einstein gas with self-interacting particles leads to
accelerated expansion of the universe. The interaction between the
particles of the gas gives naturally solutions of the type $\rho+
3p<0$, if the interaction is attractive. Phantom-accelerated
solutions with $\rho+ p<0$ can be also obtained, if we consider a
model of universe that contains the Bose gas plus radiation and in
which the interaction term initially dominates the expansion.

As such a term, responsible for the accelerated behavior,
decreases faster with the expansion than the condensate and
radiation terms, the accelerated stage eventually ends. This fact
makes the model more suitable to describe the acceleration of the
early Universe than the present accelerated era.

Our model solves the horizon problem when we
consider phantom-accelerated solutions. In addition,
nucleosynthesis is feasible by starting with a large-enough
radiation term.

%The  radiation limit  needed for
%nucleosynthesis depends on the total energy density at the
%maximum.

\emph{\textbf{Acknowledgments:}} The authors would like to
acknowledge Roc\'io Jauregui for her useful comments. G. I. also
acknowledges support from the ``Programa de Becas Postdoctorales
de la UNAM", and J. B. acknowledges  support from the Helen
program for CERN work sojourns..

\appendix

\section{}
In this Appendix we calculate a higher limit for the scale factor
$a$ for which the accelerated stage of the FRW model of
interacting gas plus radiation of Sec. II ends. Assuming that the
acceleration of the expansion given by Eq. (\ref{accB}) is
positive at the scale $a_i$, we thus  evaluate the scale factor $a$ for
which $\ddot{a}(a)=0$. This is at \be
\frac{a}{a_i}=\frac{-3\rho_{ci}+\sqrt{9\rho_{ci}^2-16\rho_{ri}v_0n^2_i}}{4\rho_{ri}}.
\label{a_0/a_i}\ee By definition, $\rho_{ri} > 0$. We define the
new variables \be y=\frac{\rho_{ci}}{\rho_{ri}}>0, \qquad
z=-\frac{v_0n^2_i/2}{\rho_{ri}}>0, \ee  and express Eq.
(\ref{a_0/a_i}) in terms of them

\be
\frac{a}{a_i}=\frac{-3y+\sqrt{9y^2+32z}}{4}.\label{a_0/a_1(y)}\ee
GR requires the total energy density to be positive
($-v_0n^2_i<\rho_{ri}+\rho_{ci}$), which reads in terms of the new
variables  $z<1+y$. Making use of the latter condition and of
Eq. (\ref{a_0/a_1(y)}), we get

\be \frac{a}{a_i}<-\frac{3y}{4}+\frac{\sqrt{9y^2+32(y+1)}}{4}.
\label{maxa_0/a_i}\ee The right-hand side term of inequality
(\ref{maxa_0/a_i}) is a decreasing function of $y$. Its maximum
value in the limit $y \rightarrow 0$ is $\sqrt{2}$. Consequently,
we conclude that the acceleration stage in the radiation plus
interacting Bose gas scenario ends at $a/a_i$ always lower than
$\sqrt{2}$.

\end{document}